\documentstyle[twoside,fleqn,espcrc2,epsfig]{article}

\newcommand{\AmS}{{\protect\the\textfont
  A\kern-.1667em\lower.5ex\hbox{M}\kern-.125emS}}

\title{Abelian dominance and adjoint color sources}

\author{Grigorios I. Poulis\address{
        National Institute for Nuclear Physics 
        and High Energy Physics (NIKHEF)\\
        P.O. Box 41882, 1009 DB Amsterdam, The Netherlands}%
        \thanks{Research supported by Human Capital and Mobility
                Grant ERBCHBICT941430.}
}
       
\begin{document}

\begin{abstract}

Abelian dominance in the case of color sources in 
the fundamental representation is shown to follow 
from certain properties of maximal abelian projected SU(2)
gauge theory. The possibility  of having an
analog of abelian dominance in the case of adjoint 
representation sources is addressed in the same framework.

\end{abstract}

\maketitle

\section{Maximal Abelian Projection revisited}

In the abelian projetion (AP) theory of confinement~\cite{tHooft}
partial gauge-fixing projects onto the U(1)$^{\rm N-1}$ Cartan 
subgroup of the original SU(N) gauge symmetry (henceforth
we take N=2). With respect to this residual abelian symmetry,
diagonal gluons transform as ``photons'', while 
off-diagonal gluons are doubly-charged matter fields.
Quark fields in the fundamental ($j$=1/2) representation
are singly-charged ($n$=1), while quarks in the
adjoint ($j$=1) representation are neutral when 
$m_j$=0 and doubly-charged ($n$=2) when $m_j$=$\pm$1. 
The maximal abelian (MA) projection~\cite{Kronfeld}, 
corresponding to maximizing
\begin{equation}
	\sum_{x,\mu}{\rm Tr}\left[
	     \sigma_3 U_{x,\mu} \sigma_3 U^\dagger_{x,\mu}\right]
	       = \sum_{x,\mu} \cos(2\phi_{x,\mu}) \ ,
	\label{ma}
\end{equation}
has provided some evidence~\cite{Suz_lat92,Poly_lat96} 
in support of this scenario of confinement. Having set
 $U_{11}$ = $\cos\phi\; e^{i\theta}$,
 $U_{12}$ = $\sin\phi\; e^{i\chi}$, etc., 
the SU(2) plaquette action $S_P$
can be decomposed as $S_P =S_\theta + S_\chi
 + S_{\theta\chi}$~\cite{Misha}, the subscripts denoting
which of the $\theta$,$\chi$ fields appear, e.g.,
$ S_\theta = (\prod_{i\in\Box} \cos\phi_i)\cos\theta_\Box$,
 where $\theta_\Box$ 
is a U(1)-invariant abelian plaquette. 
$S_\theta$ and  $S_\chi$ have one term each, while  $S_{\theta\chi}$ 
has six. Due to the gauge fixing condition,
 Eq.~(\ref{ma}), $S_\chi <  S_{\theta\chi} <S_\theta $.
Using 50 configurations on a $12^4$ lattice at $\beta=2.4$ we find 
$\langle \cos\phi\rangle \approx 2/3$ (the random value)
in local gauges~\cite{P95}. In MA projection $\langle \cos\phi\rangle $ 
is close to 1, and, accordingly, $S_\theta$ dominates the action, i.e.,
$\langle S_P\rangle$ = $\langle S_\theta\rangle$ within 9\%.
 Furthermore, we find that $\cos\phi$ 
behaves more like a parameteter than a dynamical variable and factorizes
in expectation values, e.g., $\langle \prod_{\Box}\cos\phi\rangle $
= $\langle \cos\phi\rangle ^4$ and
$\langle S_\theta\rangle $ = $\langle \prod_{\Box}\cos\phi\rangle \langle 
\cos\theta_\Box\rangle $ within 0.2\% and 3\%, respectively.
These results suggest that $\chi$ fields can be treated as beeing
essentially random, and  MAQCD, the effective abelian theory
after maximal abelian projection, is basically compact QED
with effective coupling $\beta\langle\cos\phi\rangle^4$. 

\section {Abelian Dominace}
\subsection{Operational Definition} 

A interesting feature of  MA projection, 
named ``abelian dominance'', is 
that abelian Wilson loops $W_{n=1}=\cos(\sum_{i\in L}\theta_i)$ 
reproduce the fundamental SU(2) string tension~\cite{AbelDom}.
 Abelian dominance is essentially an empirical observation. 
Operationally, one may define abelian dominance (AD) as the property
that large-scale properties of QCD are reproduced by operators $W_{\rm abel}$,
constructed exclusively from the abelian phases $\theta$~\cite{Suz_lat92}.
 We distinguish between two versions: 
\begin{itemize}
	\item \underline{\em strong version}: the operators $W_{\rm abel}$ are 
	             obtained by using (after MA projection)
	             rescaled, diagonal links,
	             $U={\rm diag}[e^{i\theta},e^{-i\theta}]$,
	    	             in the place of full SU(2) links.
	                
	\item \underline{\em weak version}: a suitable {\it Ansatz} must 
	              be devised for constructing the abelian operators.
 \end{itemize}
Clearly, if the strong version is satisfied, so is the weak, but
not {\it vice versa}.

\subsection{Fundamental Sources}

For an $T\times R$ Wilson loop in the fundamental representation, 
$W_{j=1/2}$ = $w_0+i\vec \sigma\cdot\vec w$, we can write a similar 
decomposition as for the plaquette~\cite{P95}
\begin{eqnarray}\label{w03}
w_0 + i w_3 &=& (\cos\phi)^{2L} e^{i\theta_L} + 
	      (\sin\phi)^{2L}e^{i\chi_L} \nonumber\\
   +   \sum_{m=1}^{L-1}(\cos\phi)^{2m} \!\!\!
   \!\!\!\!\!\!\!&&\!\!\!\!\!\!\!(\sin\phi)^{2(L-m)}
              \sum_{n=1}^{\scriptstyle{2L\choose 2m}} s_n 
             e^{i\Omega_n[\theta,\chi]} \ .
\end{eqnarray}
Consider now 
 $\langle  W_{j=1/2}\rangle $ = $\langle  w_0\rangle $. 
Carrying out the free-$\chi$ intregration as remarked in Sec. I, we find
\begin{equation}\label{pol1}
\langle  W_{j=1/2}\rangle \;\approx \;
\langle\cos\phi\rangle^{2L}\; \langle \cos\theta_L\rangle  \ .
\end{equation}
Since the two expectation values differ by a perimeter 
($L\equiv T+R$) term only,
they generate the same string tension. Thus, 
according to the weak version of abelian dominance, the 
abelian operator $W^{1/2}_{\rm abel}$ in this case should be 
the singly-charged abelian Wilson loop, $W_{n=1}=\cos\theta_L$. 
\begin{figure}[htb]
\begin{center}
\vskip -.7 true cm
\mbox{\epsfig{file=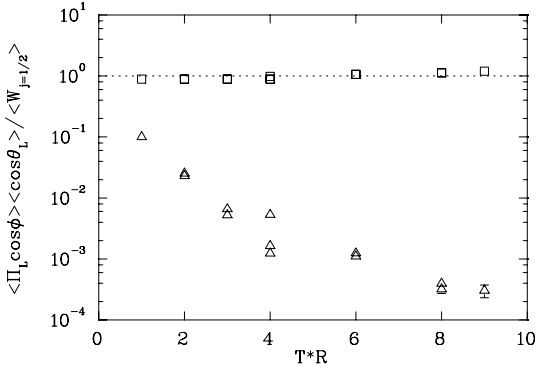,width=.30\textwidth,angle=90}}
\vskip -.6 true cm
\caption[dymmy] {Testing approximation Eq.\ (\ref{pol1}) in 
     F12 ($\triangle$) and MA ($\Box$) projection.}
\label{perc}
\vskip -.6 true cm
\end{center}
\end{figure}
\noindent
Indeed, from Fig.~\ref{perc} one sees that  Eq.~(\ref{pol1}) is a good 
approximation in MA projection 
and, not surprisingly, very bad in F12 projection. 
 Notice that $\cos\theta_L\equiv\cos(\sum_{i\in L}\theta_i)$ 
 can also be thought of as a fundamental-representation Wilson loop
 constructed from rescaled, diagonal SU(2) links $W^d_{j=1/2}$ = 
 Tr($\prod_{i\in L} {\rm diag}[e^{i\theta_i},e^{-i\theta_i}]$).
Thus, with respect to AD for fundamental repr. sources
\begin{enumerate}
\item the difference between $\cos\phi$ $\simeq$ 1 and 
      $\cos\phi$ = 1 is not essential, i.e.,
      MAQCD can be regarded as a ``diagonal SU(2)'' theory.
\item there is no distinction 
      between weak and strong versions; 
      both are satisfied.
\end{enumerate} 

\subsection{Adjoint Sources}

\begin{figure}[htb]
\begin{center}
\mbox{\epsfig{file=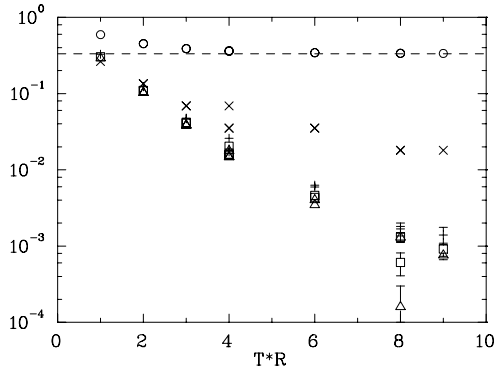,width=.30\textwidth,angle=90}}
\caption{Various adjoint Wilson loops versus loop area:
         $W_{j=1}$ ($\Box$),  $W^\pm_{j=1}$ ($\triangle$),
         $W^0_{j=1}$ (+), $W^d_{j=1}$ (o), Eq.\ \ref{exp_sq}($\times$).
         Dashed line: the limit  $W^d_{j=1}\rightarrow {1\over 3}$.
        } 
\label{test}
 \vskip -.6 true cm
\end{center}
\end{figure}

\begin{figure}[htb]
\begin{center}
\mbox{\epsfig{file=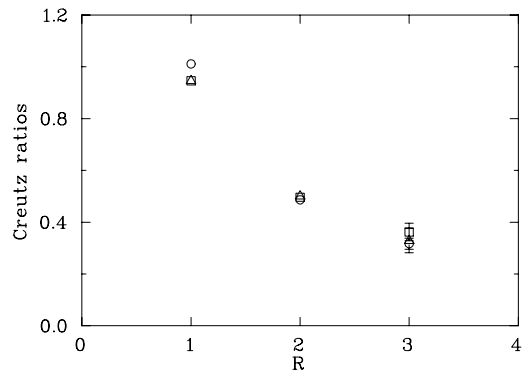,width=.30\textwidth,angle=90}}
\caption[dummy]{Creutz ratios $\chi[R,R]$ from adjoint Wilson 
loops $W_{j=1}$ (o)
        versus ones from doubly charged
        abelian loops  $W_{n=2}$ in MA projection. 
	Results from $16^4$ ($\Box$) and $12^4$ ($\triangle$) lattices.}
\label{dad}
\end{center}
 \vskip -.9 true cm
\end{figure}

Abelian dominance for adjoint sources (quarks)
requires that the adjoint ``string tension'' $\sigma_{j=1}$
can be extracted from some abelian correlator $W^1_{\rm abel}$. 
For the strong version of AD, this operator is the adjoint 
Wilson loop from diagonal SU(2) links~\cite{DelDebbio,P95}
\begin{equation}\label{strong}
     W^d_{j=1}={2\over 3} \cos(2\theta_L)+{1\over 3}  \ .
\end{equation}
      As realized by Greensite and 
coworkers~\cite{DelDebbio,Greensite}, the persistence of Casimir
 scaling ($\sigma_{j=1}\approx {8\over 3}\sigma_{j=1/2}$) 
 in MC simulations, 
and the associated failure to unambigously verify screening
for the potential between adjoint sources (\cite{Greensite}
and references therein) presents a challenge for 
abelian dominance: if MAQCD is a ``diagonal SU(2)'' theory, 
there is no way for two $m_j=0$ (neutral)
components of the adjoint source to interact via neutral
``photons'', let alone form an abelian flux tube. One would therefore expect
the adjoint string tension to vanish, rather than Casimir scale.
Indeed, if MAQCD is close to CQED, the doubly-charged abelian loop 
in Eq.~(\ref{strong}) is expected to have an area law falloff 
itself~\cite{P95,Bali} and, therefore, for large loops 
$W^d_{j=1}\rightarrow 1/3$, and the
coresponding potential vanishes. This is verified numerically
(Fig.\ \ref{test} and~\cite{P95}). Thus, the {\em strong
version of AD for adjoint sources fails}~\cite{DelDebbio,P95}.
 To see whether the weak version can be satisfied
we decompose the full adjoint Wilson loop $W_{j=1} = (4 w_0^2-1)/3$ 
into neutral ($0$) and charged ($\pm$) parts
\begin{eqnarray}
W^0_{j=1} &=& {2\over 3}(w_0^2+w_3^2)-{1\over 3}
\label{breaka}\nonumber\\
W^\pm_{j=1} &=&{2\over 3}(w_0^2-w_3^2) \ .
\end{eqnarray}
Integrating over $\chi$ as before (see ~\cite{P95} for details)
\begin{eqnarray}\label{exp_sq}
\langle w_0^2\!+\!w_3^2\rangle  \!\!\!&=&\!\!\! c^{2L} + s^{2L}
                  + \sum_{m=1}^{L-1} c^{2m} s^{2(L-m)}
                  {\scriptstyle {2L \choose 2m}}      \nonumber\\
\Rightarrow \langle W^0_{j=1}\rangle  &=& {1\over 3} \left( 1 +
 \langle\cos(2\phi)\rangle^{2L} \right)
             -{1\over 3}\ ,
\end{eqnarray}
where $c$=$\langle \cos\phi\rangle^2$, $s$=1$-c$. Thus,
without off-diagonal gluons ($\cos\phi$=1) 
$\langle W^0_{j=1}\rangle\rightarrow 1/3$, while with 
``static'' off-diagonal gluons ($\cos\phi$  $<$ 1, but fixed) 
one gets $\langle W^0_{j=1}\rangle\rightarrow 0$. 
\noindent 
However, although this greatly improves  Eq.\ (\ref{strong}), 
as seen in Fig.\ \ref{test}, 
  it is not enough for generating
 an area law for $W^0_{j=1}$: off-diagonal
 gluons are required dynamically (i.e.,
 $S_{\theta\chi}$ cannot be ignored). 
On the other hand, for the operator $(w_0^2 - w_3^2)$ 
one finds~\cite{P95} 
\begin{equation}\label{pol2}
\langle w_0^2-w_3^2\rangle \; \approx \; 
\langle \cos\phi\rangle^{4L} \; \langle\cos(2\theta_L)\rangle\ .
\end{equation}
This is the adjoint analog of Eq.~(\ref{pol1}); it is seen
~\cite{P95} to be a very good approximation in 
MA projection (only). A characteristic difference between 
$\langle W_{j=1/2} \rangle$ and $\langle W^\pm_{j=1}\rangle$ 
on one hand, and $\langle W^0_{j=1}\rangle$ on the other,
is that in the former case the free $\chi$-integration leads 
to expressions without $\langle \sin\phi\rangle$ terms, whereas in 
the latter such terms appear, with large degeneracy factors [c.f. Eq.\ 
(\ref{exp_sq})]. Since $S_{\theta\chi} << S_\theta$ relies on 
ignoring $\sin\phi$ terms, our approximations seem inconsistent 
in the case of $W^0_{j=1}$, which may explain the failure of 
generating an area law for $W^0_{j=1}$ on the basis of these
approximations. 
Numerically (200 measurements on $16^4$ and 350 on $12^4$ lattices 
at $\beta=2.4$)  we find that $W_{j=1}$,
$W^0_{j=1}$ and $W^\pm_{j=1}$ have within 10\% same 
Creutz ratios (Fig.~\ref{test}). This may be explained 
by using gauge invariance of the energy eigenstates in the spectral 
decomposition, suggesting that doubly-charged abelian Wilson loops
$W_{n=2}$ should be the relevant operators for testing {\em weak}
abelian dominance in the adjoint case. Evidence in support of
this conjecture is shown in Fig.\ \ref{dad}. Although encouraging,
this result should be treated with caution, unless some better scheme 
of approximations than the ones suggested in Sec. I succeeds in accounting
for the  area law behavior of $W^0_{j=1}$. Moreover, the pattern of weak
AD suggested here should be tested for $j=3/2$ sources as well.
Work in this direction is in progress.
\vskip -.5 true cm
 \vskip -.2 true cm

\vskip -0.1 true cm

\end{document}